\documentclass[11pt]{extarticle}
\usepackage[paperheight=12in,paperwidth=8.75in]{geometry}
\usepackage{setspace}

\usepackage{setspace}
\usepackage[T1]{fontenc}
\usepackage{times}
\usepackage{palatino}
\usepackage{lmodern}
\usepackage[latin1]{inputenc}
\usepackage{epsfig}
\usepackage[english]{babel}
\usepackage{color}
\usepackage{graphicx}
\usepackage{dcolumn}
\usepackage{amsmath,amssymb,amsfonts}
\usepackage[all]{xy}
\usepackage{cite}
\usepackage[dvipsnames]{xcolor}
\usepackage{physics}
\usepackage{mathrsfs}
\usepackage{authblk}
\usepackage{geometry}
\usepackage{abstract}
\usepackage[english]{babel}
\usepackage[T1]{fontenc}
\usepackage[colorlinks,urlcolor=blue,citecolor=blue]{hyperref}

\def\hhref#1{\href{http://arxiv.org/abs/#1}{#1}} % in bibliography

 \newcommand{\sfrac}[2]{{\textstyle\frac{#1}{#2}}}

\begin{document}
\numberwithin{equation}{section}
\newcommand{\boxedeqn}[1]{%
  \[\fbox{%
      \addtolength{\linewidth}{-2\fboxsep}%
      \addtolength{\linewidth}{-2\fboxrule}%
      \begin{minipage}{\linewidth}%
      \begin{equation}#1\end{equation}%
      \end{minipage}%
    }\]%
}

%\boxedeqn{}

\newsavebox{\fmbox}
\newenvironment{fmpage}[1]
     {\begin{lrbox}{\fmbox}\begin{minipage}{#1}}
     {\end{minipage}\end{lrbox}\fbox{\usebox{\fmbox}}}

\raggedbottom
\onecolumn

\begin{center}
{\LARGE \bf
$N$-dimensional Smorodinsky-Winternitz model and related higher rank quadratic algebra ${\cal SW}(N)$\\
}
\vspace{6mm}
{ Francisco Correa$^{a}$, Md Fazlul Hoque$^{b}$, Ian Marquette$^{c}$,  and  Yao-Zhong Zhang$^{c}$
}
\\[6mm]
\noindent ${}^a${\em 
Instituto de Ciencias F\'isicas y Matem\'aticas\\
Universidad Austral de Chile, Casilla 567, Valdivia, Chile}\\[3mm]
\noindent ${}^b${\em
Department of Mathematics \\Pabna University of Science and Technology, Pabna 6600, Bangladesh }\\[3mm]
\noindent ${}^c${\em
School of Mathematics and Physics, \\
The University of Queensland
Brisbane,  QLD 4072, Australia}\\ [1mm]
\vspace{4mm}
{\footnotesize Email: francisco.correa@uach.cl, fazlulmath@pust.ac.bd, i.marquette@uq.edu.au, yzz@maths.uq.edu.au}
\end{center}
\vskip 1cm

\begin{abstract}
\noindent The $N$-dimensional Smorodinsky-Winternitz system is a maximally superintegrable and exactly solvable model, being subject of study from different approaches. The model has been demonstrated to be multiseparable with wavefunctions given by Laguerre and Jacobi polynomials. In this paper we present the complete symmetry algebra ${\cal SW}(N)$ of the system, which it is a higher-rank quadratic one containing the recently discovered Racah algebra ${\cal R}(N)$ as subalgebra. The substructures of distinct quadratic ${\cal Q}(3)$ algebras and their related Casimirs are also studied. In this way, from the constraints on the oscillator realizations of these substructures, the energy spectrum of the $N$-dimensional Smorodinsky-Winternitz system is obtained. We show that ${\cal SW}(N)$ allows different set of substructures based on the Racah algebra ${\cal R}({ N})$ which can be applied independently to algebraically derive the spectrum of the system. 
\end{abstract}

\section{Introduction}

Superintegrable systems have attracted much attention over the years \cite{mil13} due to their very interesting properties from both mathematical and physical perspectives. The connection with the Askey-Scheme of orthogonal polynomials and quadratic algebras for all two dimensional conformally flat systems has played a clear and important role in classifying these models \cite{mil13b}. The classification of three-dimensional superintegrable systems and their symmetry algebras is still an ongoing \cite{kal1,das10} difficult problem in particular for non degenerate systems \cite{es17,cap15}. There are many families of $N$-dimensional superintegrable systems, introduced in different contexts and studied via separation of variables and factorization methods, see for examples \cite{kal02,rod02,bal11}. However, the symmetry algebras for these $N$-dimensional systems remain largely unexplored. Recently, particular classes of $N$-dimensional minimally superintegrable systems which involve higher-rank quadratic algebras with structure constants depending on Casimir operators of higher-rank Lie algebras have been discovered \cite{Hoque15,hoque15,chen19}. In general, these higher-rank quadratic algebras possess embedded structures. The derivation of these algebras is a difficult task and only few examples are known \cite{liao18,da19,gab18,ku20}. The best example is the Racah algebra ${\cal R}(N)$ \cite{gab18,ku20} for which different realization and presentation exist. The relationship with the Howe duality, co-product ${\cal U}(su(1,1))^{\otimes N}$, Temperley-Lieb algebra, Brauer algebras and Racah polynomials has been discussed in \cite{bie20}. Recent works \cite{lat21a,lat21b} on the coalgebra approach demonstrated how partial Casimirs provide integrals of motion for a wide range of superintegrable models which close to a quadratic algebra with higher-order Serre-type relations. There, the connection with the Racah algebra ${\cal R}(N)$ as well as chains of ${\cal R}(3)$ was also established. 

The Smorodinsky-Winternitz (SW)\footnote{We dedicate this paper to the memory of Prof. Pavel Winternitz, for the numerous discussions on different topics of mathematical physics and his inspiration in the search of symmetries in physics.} systems on two- and three-dimensional Euclidean spaces \cite{win67,mak67,ev90} are examples of maximally superintegrable systems. It was shown how the R-matrix approach can be applied \cite{gag85} to the Rosochatius model which is a generalization of the SW system. The connection with the Gaudin magnetic system was presented in \cite{gag85} and for the 2D case the interpretation as one of the 2D Krall-Scheffer operators was done in \cite{har01}. 
While the quadratic algebras and algebraic derivation of their spectra were presented in \cite{das01},  a similar approach was studied for other 3-dimensional models  in \cite{post11}. $N$-dimensional analogs of the SW system have been formulated in \cite{ev91,ev90b}. 

The purpose of this paper is to give an algebraic derivation of the spectrum of the $N$-dimensional SW system based on the complete symmetry algebra. We present various sub-algebraic structures of the symmetry algebra which consist of distinct quadratic algebras ${\cal Q}(3)$ and their Casimirs, and show how these substructures enable us to algebraically determine the spectrum of the SW system. 

This paper is organized in the following way, in Section 2 we present the solution of the SW model via separation of variables in cartesian and hyperspherical coordinates. In Section 3, we present the integrals and the corresponding higher-rank quadratic algebra ${\cal SW}({N})$ of the model. In Section 4 we introduce the notion of substructures for this model related to quadratic algebra ${\cal Q}(3)$ \cite{das01,gra92} and present an algebraic derivation of the spectrum. In Section 5 we give an algebraic derivation of spectrum based on the Racah algebra $ {\cal R}({ N})$. Section 6 provides the conclusion.

\section{Separation of variables}

The $N$-dimensional Smorodinsky-Winternitz Hamiltonian operator is
\begin{equation}
 H= -\frac{1}{2} \sum_{i=1}^N \partial_i^2 + b  \sum_{i=1}^N  x_i^2 +  \sum_{i=1}^N \frac{a_i}{x_i^2},\label{H1}
\end{equation}
where all masses are equal and we set $\hbar=m_i=1$,  $\partial_i={\partial}/{\partial x_i}$. 
Let us recall some known facts on this model that has been studied extensively. The corresponding Schr\"{o}dinger equation and Hamilton-Jacobi equation allow separation of variables in various coordinate systems. In quantum mechanics, the separation of variables in Cartesian coordinates of the $H \Psi =E \Psi$ \cite{ev91} is done via
\begin{equation}
\Psi=\prod_{i=1}^{N}\psi_{n_i}\ , \quad 
\psi_{n_i}=N_{n_i} e^{-\sqrt{\frac{b}{2}} x_i^2} x_i^{\frac{1}{2} \pm \nu_i} L_{n_i}^{\pm \nu_i}(\sqrt{2b} x_i^2),
\end{equation}
in terms of the associated Laguerre polynomials $L_{n}^{a}(x)$. Here we set  the constants
\begin{equation}
\nu_i =\frac{1}{2} \sqrt{1{+}8 a_i} \, , \quad i=1,...,N
\end{equation}
and the normalization ones as
\begin{equation}
N_{n_i}=(2b)^{\frac{1}{4}} \left[  (2b)^{\pm \nu_i} { \Gamma(n_i +1)}/ {\Gamma(n_i +1 \pm \nu_i)} \right]^{\frac{1}{2}}.
\end{equation}
The corresponding spectrum and  the degeneracies of each level are 
\begin{equation}
E=  \sqrt{2 b}  \sum_{i=1}^{N}(2n_i  \pm\nu_i+ 1), \qquad {\rm deg}(n)=\begin{pmatrix}
N+n-1 \\
N-1
\end{pmatrix}, \label{En1}
\end{equation}
where $n=\sum_{i=1}^{N}n_i$.

%
%The $N$-dimensional Smorodinsky-Winternitz model Hamiltonian is given by
%\begin{eqnarray}
% H= -\frac{1}{2} \sum_{i} \partial_i^2 + b \sum_i x_i^2 + \sum_i\frac{a_i}{x_i^2}.\label{H1}
%\end{eqnarray}
%Let us recall some known facts on this model. The corresponding Schr\"{o}dinger equation and Hamilton-Jacobi equation allow separation of variables in various coordinate systems. In the Cartesian coordinates separation of variables gives rise to the wave functions
%\begin{eqnarray}
%\psi_{n_i}(x_i)=N_{n_i} e^{-\sqrt{\frac{b}{2}} x_i^2} x_i^{\frac{1}{2} \pm \nu_i} L_{n_i}^{\pm \nu_i}(\sqrt{2b} x_i^2),
%\end{eqnarray}
%where $n_i$ are positive integers, $\nu_i =\frac{1}{2} \sqrt{1+ 8 a_i}$,
%\begin{eqnarray}
%N_{n_i}=(2b)^{\frac{1}{4}} \left[  (2b)^{\pm \nu_i} \frac{ \Gamma(n_i +1)}{\Gamma(n_i +1 \pm \frac{1}{2} \sqrt{1+8 a_i})} \right]^{\frac{1}{2}},\nonumber
%\end{eqnarray}
%and the corresponding spectrum 
%\begin{eqnarray}
%E=  \sqrt{2 b} \left[ \sum_{i=1}^{N}\left(2n_i  \pm \frac{1}{2}\sqrt{ 1+ 8 a_i}\right) + N\right].\label{En1}
%\end{eqnarray}
%The degeneracy of each energy level is 
%\begin{eqnarray*}
%g(n)=\begin{pmatrix}
%N+n-1 \\
%N-1
%\end{pmatrix},~~~~~~ n=\sum_{i=1}^{N}n_i.
%\end{eqnarray*}
%

\subsection{Hyperspherical coordinates}

The system (\ref{H1}) is also separable in $N$-dimensional hyperspherical coordinates, defined  by
\begin{eqnarray}
&x_{1}&=r\cos\theta_1 ,
\nonumber\\& x_{2}&=r \sin\theta_1\cos\theta_2,
\nonumber\\& x_{3}&=r \sin\theta_1\sin \theta_2 \sin \theta_3,
\nonumber\\&...&
\nonumber\\&  x_{N-1}&=r\sin\theta_1\sin\theta_2\dots \sin\theta_{N-2}\cos\theta_{N-1},
\nonumber\\&  x_{N}&=r\sin\theta_1\sin\theta_2\dots \sin\theta_{N-2}\sin\theta_{N-1}.\label{kfpf1}
\end{eqnarray}
The $N$ $x_{i}$'s are the cartesian coordinates in terms of the hyperspherical angles $\{\theta_1, \dots, \theta_{N-1}\}$ and radius $r$. Thus, the Hamitonian operator (\ref{H1}) reduces to
%\begin{eqnarray}
%H &&= -\frac{1}{2}\left[\frac{\partial^2}{\partial r^2}+\frac{N-1}{r}\frac{\partial}{\partial r} -2br^2 \right]-\frac{1}{2r^2}\left\{\left(\frac{\partial^2}{\partial \theta^2_1}+(N-2)\cot\theta_1\frac{\partial}{\partial\theta_1}-\frac{2a_1}{\cos^2\theta_1}\right.\right. \nonumber \\&& \left.
% +\frac{1}{\sin^2\theta_1}\left(\frac{\partial^2}{\partial \theta^2_2}+(N-3)\cot\theta_2\frac{\partial}{\partial\theta_2}-\frac{2a_2}{\cos^2\theta_2}\right.\right. \nonumber \\&& \left.
%  +\frac{1}{\sin^2\theta_2}\left(\frac{\partial^2}{\partial \theta^2_3}+(N-4)\cot\theta_3\frac{\partial}{\partial\theta_3}-\frac{2a_3}{\cos^2\theta_3}\right.
%  \right. \nonumber \\&& \left.\qquad \dots\right. \nonumber \\&& \left. \qquad\dots \right. \nonumber \\&& \left.\left.\left.\left. 
%+\frac{1}{\sin^2\theta_{N-3}}\left(\frac{\partial^2}{\partial \theta^2_{N-2}}+\cot\theta_{N-2}\frac{\partial}{\partial\theta_{N-2}}-\frac{2a_{N-2}}{\cos^2\theta_{N-2}}\right)...\right)\right)\right)
%\right. \nonumber \\&& \left. 
%+\frac{1}{\sin^2\theta_{N-2}}\left(\frac{\partial^2}{\partial \theta^2_{N-1}}-\frac{2a_{N-1}}{\cos^2\theta_{N-1}}-\frac{2a_{N}}{\sin^2\theta_{N-1}}\right)\right\}.
%\end{eqnarray}
\begin{align}
H &= -\frac{1}{2}\left[\frac{\partial^2}{\partial r^2}+\frac{N-1}{r}\frac{\partial}{\partial r} -2br^2 \right]-\frac{1}{2r^2}\left\{\left(\frac{\partial^2}{\partial \theta^2_1}+(N-2)\cot\theta_1\frac{\partial}{\partial\theta_1}-\frac{2a_1}{\cos^2\theta_1}\right.\right. \nonumber \\& \left.
 +\frac{1}{\sin^2\theta_1}\left(\frac{\partial^2}{\partial \theta^2_2}+(N-3)\cot\theta_2\frac{\partial}{\partial\theta_2}-\frac{2a_2}{\cos^2\theta_2}\right.\right. \nonumber \\& \left.
  +\frac{1}{\sin^2\theta_2}\left(\frac{\partial^2}{\partial \theta^2_3}+(N-4)\cot\theta_3\frac{\partial}{\partial\theta_3}-\frac{2a_3}{\cos^2\theta_3}\right.
  \right. \nonumber \\& \left.\qquad \dots\right. \nonumber \\& \left. \qquad\dots \right. \nonumber \\& \left.
+\frac{1}{\sin^2\theta_{N-3}}\left(\frac{\partial^2}{\partial \theta^2_{N-2}}+\cot\theta_{N-2}\frac{\partial}{\partial\theta_{N-2}}-\frac{2a_{N-2}}{\cos^2\theta_{N-2}}\right.
\right. \nonumber \\& \left. \left.\left.\left. \left. 
+\frac{1}{\sin^2\theta_{N-2}}\left(\frac{\partial^2}{\partial \theta^2_{N-1}}-\frac{2a_{N-1}}{\cos^2\theta_{N-1}}-\frac{2a_{N}}{\sin^2\theta_{N-1}}\right)\right)...\right)\right)\right)\right\}.
\end{align}
The ansatz 
\begin{equation}
\Psi=\psi(r)\prod_{l=1}^{N-1}\psi(\theta_l)
\end{equation}
in the Schr\"{o}dinger equation $H\Psi=E\Psi$, reduce to the following equations
\begin{align}
-\frac{1}{2}\left(\frac{\partial^2}{\partial r^2}+\frac{N-1}{r}\frac{\partial}{\partial r} -2br^2 +\frac{k_1}{r^2}\right)\psi(r)&=E\psi(r), \label{An0}
  \\
\left(\frac{\partial^2}{\partial \theta^2_{\ell}}+(N-\ell-1)\cot\theta_{\ell}\frac{\partial}{\partial\theta_{\ell}}-\frac{2a_{\ell}}{\cos^2\theta_{\ell}}+\frac{k_{\ell+1}}{\sin^2\theta_{\ell}} \right)\psi(\theta_\ell)&=-k_\ell\psi(\theta_\ell), \label{An1}
 \\
\left(\frac{\partial^2}{\partial \theta^2_{N-1}}-\frac{2a_{N-1}}{\cos^2\theta_{N-1}}-\frac{2a_{N}}{\sin^2\theta_{N-1}}\right)\psi(\theta_{N-1})&=-k_{N-1}\psi(\theta_{N-1}),\label{An2}
\end{align}
where $\ell=1,2,\dots, N-2$. In order to solve these equations,  we can observe first that (\ref{An2}) can be converted, by choosing $z=\sin^2(\theta_{N-1})$ and then $\psi(z)=z^\alpha(1-z)^\beta f(z)$, to the form
\begin{equation}
 z(1{-}z)f''(z)+\left\{2\alpha{+}\frac{1}{2}{-}(2\alpha{+}2\beta{+}1)z\right\}f'(z) +\left(\frac{k_{N-1}}{4}{-}(\alpha{+}\beta)^2\right) f(z)=0,\label{An4}
\end{equation}
where $2\alpha=\pm \nu_N+\frac{1}{2}$, $2\beta=\pm\nu_{N-1}+\frac{1}{2}$. The equation (\ref{An4}) is nothing else that the  Jacobi equation
\begin{equation}
x(1{-}x)y''+\left(\lambda-(\omega+1)x\right)y'+\tau_{N-1}(\tau_{N-1}+\omega) y=0,\label{Hy1}
\end{equation}
with coefficients $\omega=1\pm \nu_N\pm \nu_{N-1}$ and $k_{N-1}=(2\tau_{N-1}+1\pm \nu_N \pm  \nu_{N-1})^2$.
 We can write the solution in terms of Jacobi polynomials for the angular part,
 \begin{equation}%\label{}
\psi(\theta_{N-1})\varpropto\cos^{1/2\pm \nu_{N-1}}(\theta_{N-1}) \sin^{1/2\pm \nu_N}(\theta_{N-1}) P^{(\pm\nu_{N}, \pm\nu_{N-1})}_{\tau_{N-1}}(\cos(2\theta_{N-1})).
\end{equation}
If we consider $\ell=N-2$ in Eq. (\ref{An1}) we can set a similar substitution from above $z=\sin^2(\theta_{N-2})$ and then $\psi(z)=z^{\alpha_1}(1{-}z)^{\beta_1}f(z)$, to
 \begin{equation}%\label{}
z(1{-}z)f''{+}\left\{2\alpha_1{+}1{-}\left(2\alpha_1{+}2\beta_1{+}\frac{3}{2}\right)z\right\}f'{+}\left(\frac{k_{N-2}}{4}{-}[\alpha_1{+}\beta_1]\left[\alpha_1{+}\beta_1{+}\frac{1}{2}\right]\right) f=0,\label{Ann4}
\end{equation}
where $2\alpha_1=2\tau_{N-1}+1\pm \nu_N \pm \nu_{N-1}$, $2\beta_1=\pm\nu_{N-2}+\frac{1}{2}$. Now comparing the above equation with (\ref{Hy1}), we find the identifications
\begin{align}
 \alpha&=2\tau_{N-1}+\frac{3}{2}\pm \nu_N \pm \nu_{N-1}\pm \nu_{N-2}, 
\\
k_{N-2}&=(2\tau_{N-1}+2\tau_{N-2}+2\pm \nu_N \pm  \nu_{N-1}\pm  \nu_{N-2})^2-\frac{1}{4}
\end{align}
and the solutions in terms of the Jacobi polynomials as
\begin{eqnarray}
&&\psi(\theta_{N-2})\varpropto \cos^{1/2\pm \nu_{N-2}}(\theta_{N-2}) \sin^{2\tau_{N-1}+1\pm \nu_N \pm \nu_{N-1}}(\theta_{N-1})\nonumber\\&&\qquad\qquad\quad\times P^{(2\tau_{N-1}+1\pm \nu_N \pm \nu_{N-1}, \pm\nu_{N-2})}_{\tau_{N-2}}(\cos(2\theta_{N-2})).
\end{eqnarray}
The same strategy is used for other $l$ values in (\ref{An1})  yielding 
\begin{equation}
\psi(\theta_{l})\varpropto \cos^{1/2\pm \nu_{l}}(\theta_{l}) \sin^{\mu_{l+1}+1-(N-l)/2}(\theta_{N-1}) P^{(\mu_{l+1}, \pm\nu_{l})}_{\tau_{l}}(\cos(2\theta_{l})),
\end{equation}
where the separations constants $k_l$ and parameters $\mu_l$ take the form
\begin{align}
k_l&=\left[2\sum_{i=l}^{N-1} \tau_{i}\pm \sum_{i=l}^{N} \nu_i +(N-l)\right]^2-\frac{1}{4}(N-l-1)^2, \quad l=1,\dots,N-3,  \\
\mu_l &=2\sum_{i=l}^{N-1} \tau_{i}\pm \sum_{i=l}^{N} \nu_i+\frac{N-l-2}{2}, \quad l=1,\dots,N-1. 
\end{align}
We now calculate $\psi(r)$ for the radial equation (\ref{An0}).
By setting  
 $z=\varepsilon r^2$, $R(z)=z^{\nu-\frac{N-2}{4}} e^{-\frac{z}{2}}f(z)$ and using the values of the separation constant $k_1$, we can convert the radial equation into the form
\begin{equation}
z\frac{d^2f(z)}{dz^2}+(2\nu+1-z)\frac{df(z)}{dz}+\left[\frac{E}{2\varepsilon}-\nu-\frac{1}{2}\right]f(z)=0, \label{an6}
\end{equation}
where
\begin{equation}
2\nu=2\sum_{i=1}^{N-1} \tau_{i}\pm \sum_{i=1}^{N} \nu_i +(N-1),\quad \varepsilon=\sqrt{2b}.\label{La1}
\end{equation}
Setting 
\begin{equation}
\tau_r=\frac{E}{2\varepsilon}-\nu-\frac{1}{2},\qquad \text{where}\,\,  \tau_r\,\,  \text{are positive integers}, \label{tau1}
\end{equation}
then (\ref{an6}) can be identified with the associated Laguerre differential equation and the solution acquires the form
%\begin{equation}\label{kl}
%\psi(r)\propto\frac{(\varepsilon r^2)^{\nu-\frac{N-2}{2}}e^{-\frac{\varepsilon r^2}{2}}}{\sqrt{2\nu+1}}
%{}_1F_1(\tau_r,2\nu+1,\varepsilon r^2).
%\end{equation} 
\begin{equation}\label{kl}
\psi(r):=\psi_{\tau_r}^{2\nu}(r) \propto e^{-\sqrt{\frac{b}{2}}r^2}r^{2\nu-\frac{N-2}{2}}L^{2\nu}_{\tau_r}(\varepsilon r^2).
\end{equation}
Now from (\ref{tau1}) and using the values of $\varepsilon$ and $\nu$, we obtain the energy spectrum of the system,
\begin{equation}
E=\sqrt{2b}\left(2\tau_r+2\sum_{i=1}^{N-1} \tau_{i}\pm \sum_{i=1}^{N} \nu_i +N\right).\label{En2}
\end{equation}
This expression can be related to the dimensions of the irreducible representations of the Lie algebra $su(N)$. However, the symmetry algebra of the system is not a Lie algebra, even in two dimensions. The various algebraic descriptions of superintegrable systems usually rely on a priori knowledge of the wavefunctions, and then using properties of the related orthogonal polynomials to construct other type of operators. Our approach is different, we consider only the explicit differential operator realizations of the Hamiltonian and related integrals to derive its spectrum. It is necessary to construct explicitly the underlying symmetry algebra in order to present this algebraic derivation. However, the symmetry algebra of the Hamiltonian (\ref{H1}) is a higher rank quadratic algebra, which is by far more complicated structure than Lie algebras containing less studied algebraic structures.
\section{Superintegrability and symmetry algebra ${\cal SW}({ N})$}
The $N$-dimensional Smorodinsky-Winternitz system (\ref{H1}) is superintegrable. It has the following second order integrals of motion
\begin{eqnarray}\label{thebs}
&& B_i= - \partial_i^2 + 2 bx_i^2+ 2 \frac{a_i}{x_i^2} , \\&& \label{theas}
  A_{ij}= - J_{ij}^2  + 2 \frac{a_i x_j^2}{x_i^2}+ 2 \frac{a_j x_i^2}{x_j^2} +\frac{1}{2}\quad(=A_{ji}),
\end{eqnarray}
where
\begin{eqnarray}
   J_{ij}= x_i  \partial_j- x_j \partial_i,\quad i, j=1,2,\dots, N. 
\end{eqnarray}
From the definition of the Hamiltonian (\ref{H1}), it is clear the integrals $B_i$ satisfy
\begin{equation}\label{sumb}
H=\frac{1}{2} \sum_i^{N}B_i.
\end{equation}
We can easily verify the following commutation relations
\begin{eqnarray}
[H,B_i]= [H,A_{ij}]= [B_i,B_j]=[A_{ij}, B_k]=0, \quad i,j, k=1,2,\dots, N\quad \text{and}\quad k\neq i, j.
\end{eqnarray}
For later convenience, we present a diagram representation of the above relations.
\begin{eqnarray}
\begin{xy}
(0,0)*+{A_{ij}}="m"; (15,0)*+{B_k}="q"; (30,0)*+{B_i}="p"; (45,0)*+{A_{jk}}="n"; (20,20)*+{H}="r"; 
"r";"p"**\dir{--};
"m";"q"**\dir{--}; 
"n";"p"**\dir{--};
"q";"p"**\dir{--}; 
"n";"r"**\dir{--};
"r";"m"**\dir{--};
"r";"q"**\dir{--};
\end{xy}
\quad
\begin{xy}
(0,0)*+{A_{ik}}="m"; (15,0)*+{B_j}="q"; (30,0)*+{B_i}="p"; (45,0)*+{A_{jk}}="n"; (20,20)*+{H}="r"; 
"r";"p"**\dir{--};
"m";"q"**\dir{--}; 
"n";"p"**\dir{--};
"q";"p"**\dir{--}; 
"n";"r"**\dir{--};
"r";"m"**\dir{--};
"r";"q"**\dir{--};
\end{xy}
\end{eqnarray}
We can further define more conserved charges
\begin{equation}
 C_{ij}= [B_i,A_{ij}]= [B_j,A_{ij}], \label{E1.6}
\quad
 D_{ijk}=[A_{ij},A_{jk}],
\end{equation}
\begin{equation}
 [C_{ij},H]=0=[D_{ijk},H]. 
\end{equation}
It can be shown that the above constants of motion of the system (\ref{H1}) close to satisfy the following quadratic symmetry algebra ${\cal SW}({ N})$ relations,
\begin{eqnarray}
&&[ A_{jk},D_{ijk}] =  4 \{A_{ik},A_{jk}\}{-} 4 \{A_{jk},A_{ij}\}{+} 4(8 a_j{-}3) A_{ik}{-}4(8 a_k{-}3) A_{ij}, \\&&
[ A_{kl},D_{ijk}]  =  4 \{A_{ik},A_{jl}\}{-} 4 \{A_{jk},A_{il}\},\\ &&
 [D_{ijk},D_{jkl}] =4 \{D_{jkl},A_{ij}\}{-}4 \{D_{ikl},A_{jk}\} 
{ -}4 \{D_{ijk},A_{jl}\}{-}4(8 a_j{-}3)D_{ikl}, \\ &&
 [D_{ijk},D_{klm}] =4 \{D_{ilm},A_{jk}\}{-}4 \{D_{jlm},A_{ik}\}, \\ &&
 [C_{ik},C_{kl}] =4 \{C_{li},B_k\}, \\ &&
 [B_{i},D_{ijk}] =4 \{B_k,A_{ij}\}{-}4 \{B_j,A_{ik}\}, \\ &&
 [B_i,C_{ij}] =-4 \{B_i,B_j\} {+}32 b A_{ij}, \\&&
  [ C_{ij},D_{jkl}]=4\{C_{il},A_{jk}\}{-}4 \{C_{ik},A_{jl}\},   \\ &&
  [ C_{ij},D_{ijk}]= -4 \{C_{ik} A_{ij}\}{-}4 \{C_{jk},A_{ij}\}, \\ &&
  [A_{ij},C_{ij}]=4\{A_{ij},B_{j}\}{ -}4 \{A_{ij},B_{i}\}{-} 4(8 a_j{-}3)B_i {+} 4(8 a_i{-}3)B_j, \\ &&
   [A_{ij},C_{ki}]=4\{A_{kj},B_{i}\}{ -}4 \{A_{ik},B_{j}\} , \end{eqnarray}
where $ i \neq j \neq k \neq l \neq m$ with  $i,j,k,l,m \in \{1,...,N\}$ covering all non-vanishing commutators. The relations involving $A_{ij}$ and $D_{lmn}$ define the Racah algebra ${\cal R}({N})$, which has been the subject of attention in last years with connections to many other algebraic structures. It is interesting to see ${\cal R}({N})$ is embedded in the larger symmetry algebra ${\cal SW}({N})$ of the $N$-dimensional Smorodinsky-Winternitz system.

\section{Subalgebra structures and energy spectrum}

\subsection{ The algebra ${\cal Q}(3)$ }  \label{subs4}

The structures of the algebras ${\cal SW}({N})$ and ${\cal R}({N})$ are complicated for $N>3$ as they are then of higher rank. In order to algebraically deriving the spectrum we exploit the existence of set of commutating integrals as well as the existence of different subalgebras, referred to as substructures, involving 3 generators. Each of these substructures has similarity with the quadratic algebra ${\cal Q}(3)$ introduced in context of two-dimensional systems  \cite{das01}. The algebraic approach involves identifying $N$ substructures (${\cal Q}_i(3)$ \cite{liao18,da19 }, each involving 3 generators
 $\{E_i, F_i, G_i\}$ for any {\em fixed} $i=1,...N$. They satisfy the following general commutation relations
\begin{eqnarray}
[E_i,F_i]&=&G_i,\nonumber\\
{}[E_i,G_i]&=&\alpha_i A_i^2+\gamma_i \{E_i,G_i\}+\delta_i E_i +\epsilon_i F_i +\zeta_i,\nonumber\\
{}[F_i,G_i]&=&a_i E_i^2-\gamma_i F_i^2-\alpha_i \{E_i,F_i\} +d_i E_i-\delta_i F_i+z_i,    \label{E19}
\end{eqnarray}
where $i$ is a fixed value in $i=1,...,N$ and $[E_i,E_j]=0$, $\forall \quad i,j$. The structure constants for each of the substructures, $\alpha_i$, $\gamma_i$, $\delta_i$, $\epsilon_i$, $\zeta_i$, $a_i$, $d_i$, $z_i$, are constants or more generally polynomials of central elements of the $i$-th substructure, e.g. the Hamiltonian $H$ and the generators of the other substructures which commute with the generators of the $i$-th substructure. Each of the substructures has a Casimir invariant, which is cubic in the generators $\{E_i,F_i,G_i\}$ of the given substructure
\begin{align}
K_i&=G_i^{2}-\alpha_i\{E_i^2,F_i\}-\gamma_i \{E_i,F_i^{2}\}+(\alpha_i \gamma_i-\delta_i)\{E_i,F_i\}+ (\gamma_i^2-\epsilon_i) F_i^{2} \nonumber\\
&+ (\gamma_i\delta_i-2\zeta_i) F_i+\frac{2a_i}{3}E_i^3 +\left(d_i+\frac{a_i\gamma_i}{3}+\alpha_i^2\right) E_i^{2} +\left(\frac{a_i\epsilon_i}{3}+\alpha_i\delta_i+2z_i\right) E_i . \label{E20}
\end{align}
This Casimir can also be realized purely in terms of the Hamiltonian and central elements of the  substructure. Then as shown in \cite{das01}, the quadratic algebra ${\cal Q}(3)$ (\ref{E19}) for any fixed $i$ value can be realized in terms of the deformed oscillator algebra,
\begin{equation}
[\aleph_i,b_i^{\dagger}]=b_i^{\dagger},\quad [\aleph_i,b_i]=-b_i,\quad b_ib_i^{\dagger}=\Phi (\aleph_i+1),\quad b_i^{\dagger} b_i=\Phi(\aleph_i). \label{Deform}
\end{equation}
The structure function is given by
\begin{align}
 \Phi_i(n_i)  &=  \frac{1}{4}\left[{-}\frac{K_i'}{\epsilon_i}{-}\frac{z_i}{\sqrt{\epsilon_i}}{-}\frac{\delta_i}{\sqrt{\epsilon_i}}\frac{\zeta_i}{\epsilon_i}{+}\left(\frac{\zeta_i}{\epsilon_i}\right)^2\right]  \nonumber\\
 & -\frac{1}{12}\left[3d_i{-}a_i\sqrt{\epsilon_i}{-}3\alpha_i \frac{\delta_i}{\sqrt{\epsilon_i}}{+}3 \frac{\delta_i^2}{\epsilon_i}{-}6\frac{z_i}{\sqrt{\epsilon_i}}{+}6\alpha_i\frac{\zeta_i}{\epsilon_i}-6\frac{\delta_i}{\sqrt{\epsilon_i}}\frac{\zeta_i}{\epsilon_i}\right](n_i{+}u_i)\nonumber\\
&+\frac{1}{4}\left[ \alpha_i^2{+}d_i{-}a_i\sqrt{\epsilon_i}{-}3\alpha_i\frac{\delta_i}{\sqrt{\epsilon_i}}{+}\frac{\delta_i^2}{\epsilon_i}{+}2\alpha_i\frac{\zeta_i}{\epsilon_i}  \right](n_i{+}u_i)^2
 \nonumber\\
 &-\frac{1}{6}\left[3\alpha_i^2{-}a_i\sqrt{\epsilon_i}{-}3\alpha_i\frac{\delta_i}{\sqrt{\epsilon_i}}  \right](n_i{+}u_i)^3{+}\frac{1}{4}\alpha^2(n_i{+}u_i)^4
\end{align}
for $\gamma_i=0$, $\epsilon_i \neq 0$, and by 
%\begin{eqnarray}
% \Phi(n_i) &=& 768 (\alpha_i  \epsilon_i^2+4 \gamma_i^2 \zeta_i -2 \gamma_i  \delta_i  \epsilon_i )^2+32 \gamma_i^4 (2 (n_i+u_i)-1)^2 \nonumber \\ 
%& &
%\left(12 (n_i+u_i)^2 -12 (n_i+u_i)-1\right) \left(3 \alpha_i^2 \epsilon_i^2+4 \alpha_i  \gamma_i^2 \zeta_i -6 \alpha_i  \gamma_i  \delta_i  \epsilon_i +2 a \gamma_i  \epsilon_i^2 \right.\nonumber \\
%& & \left.
%+2 \gamma_i^2 \delta_i^2-4 \gamma_i^2 d \epsilon_i +8 \gamma_i^3 z_i -48 \gamma_i^6\right) \left(2 (n_i+u_i)-3) (2 (n_i+u_i)-1)^4 \right) \nonumber  \\
%& &
% \left(2 (n_i+u_i)+1\right)^2 (\alpha_i^2 \epsilon_i -\alpha_i  \gamma_i  \delta_i + a_i \gamma_i  \epsilon_i -\gamma_i^2 d_i)-256 \gamma_i^2 \left(2 (n+i+u_i)-1\right)^2 \nonumber  \\
%& &
%(3 \alpha_i^2 \epsilon_i^3+4 \alpha_i  \gamma_i^4 \zeta_i +12 \alpha_i  \gamma_i^2 \zeta_i  \epsilon_i -9 \alpha_i  \gamma_i  \delta_i  \epsilon_i^2+a_i \gamma_i  \epsilon_i^3+2 \gamma_i^4 \delta_i^2 \nonumber \\
%& &
%-12 \gamma_i^3 \delta_i  \zeta_i +6 \gamma_i^2 \delta_i^2 \epsilon_i +2 \gamma_i^4 d_i \epsilon_i -3 \gamma_i^2 d_i \epsilon_i^2-4 \gamma_i^5 z+12 \gamma_i^3 z_i \epsilon_i )  \nonumber  \\
%& &
%+ \gamma_i^8 \left (2 (n_i+u_i)-3\right)^2 \left(2 (n_i+u_i)-1\right)^4 \left(2 (n_i+u_i)+1\right)^2 (3 \alpha_i^2 + 4 a_i \gamma_i ) \nonumber \\
%& &
%-3072 \gamma_i^6  K_i\left(2 (n_i+u_i)-1\right )^2 
%\end{eqnarray}
\begin{align}\nonumber
 \Phi_i(n_i) &=\gamma_i^8 (3 \alpha_i^2 {+ }4 a_i \gamma_i )[2 (n_i{+}u_i){-}3]^2 [2 (n_i{+}u_i){-}1]^4 [2 (n_i{+}u_i){+}1]^2-3072 \gamma_i^6  K_i[2 (n_i{+}u_i){-}1]^2    \\ \nonumber
&-48  \gamma_i^6(\alpha_i^2 \epsilon_i{-}\alpha_i  \gamma_i  \delta_i{+} a_i \gamma_i  \epsilon_i{-}\gamma_i^2 d_i)[2 (n_i{+}u_i){-}1]^4[2 (n_i{+}u_i){+}1]^2[2 (n_i{+}u_i){-}3] \\
&+32 \gamma_i^4  \left(3 \alpha_i^2 \epsilon_i^2{+}4 \alpha_i  \gamma_i^2 \zeta_i {-}6 \alpha_i  \gamma_i  \delta_i  \epsilon_i {+}2 a_i \gamma_i  \epsilon_i^2{+}2 \gamma_i^2 \delta_i^2{-}4 \gamma_i^2 d_i \epsilon_i{+}8 \gamma_i^3 z_i\right) \times \\ \nonumber
&[2 (n_i{+}u_i){-}1]^2[12 (n_i{+}u_i)^2{-}12 (n_i{+}u_i){-}1]+ 768 (\alpha_i  \epsilon_i^2+4 \gamma_i^2 \zeta_i -2 \gamma_i  \delta_i  \epsilon_i )^2 \\ \nonumber
&-256\gamma_i^2[2 (n_i{+}u_i){-}1]^2(3 \alpha_i^2 \epsilon_i^3{+}4 \alpha_i  \gamma_i^4 \zeta_i {+}12 \alpha_i  \gamma_i^2 \zeta_i  \epsilon_i{-}9 \alpha_i  \gamma_i  \delta_i  \epsilon_i^2{+}a_i \gamma_i  \epsilon_i^3{+}2 \gamma_i^4 \delta_i^2 \\ \nonumber
&-12 \gamma_i^3 \delta_i  \zeta_i{+}6 \gamma_i^2 \delta_i^2 \epsilon_i{+}2 \gamma_i^4 d_i \epsilon_i{-}3 \gamma_i^2 d_i \epsilon_i^2{-}4 \gamma_i^5 z_i{+}12 \gamma_i^3 z_i \epsilon_i )
\end{align}
for $\gamma_i \neq 0$. The structure function $\Phi_i(n_i)$ depends on the Hamiltonian $H$, a constant $u_i$, the eigenvalues $n_i$  and the ones associated with the number operators of the given ${\cal Q}_i(3)$ and the central operators $E_i$, respectively. In this way the structure functions involve eigenvalues of mutually commuting operators. The construction of the deformed oscillator algebra \cite{das01} rely on the integrals $E_i$ being realized only in terms of the number operators ${\cal N}_i$ associated with $n_i$. They provide constraints for the eigenvalues of the operator $E_i$ 
\begin{eqnarray}
&&e(E_i)=E_i(q_i)=\frac{\gamma_i}{2}\left((q_i+u_i)^2-\frac{\epsilon_i}{\gamma_i^2}-\frac{1}{4}\right), \qquad \gamma_i \neq 0;\label{E3.21}
 \\&&
e(E_i)=E_i(q_i)=\sqrt{\epsilon_i}(q_i+u_i), \qquad \gamma_i=0,\quad \epsilon_i \neq 0.\label{E3.22}
\end{eqnarray} 
Here we denote the eigenvalues of the generators $E_i$ in terms of $q_i$. Other constraints on the structure functions $\Phi_i(n_i,u_i,H)$ of each substructures take the form of $\Phi_i(0,u_i,H)=0$ and $\Phi_i(p_i{+}1,u_i,H)=0$ where $q_i=0,1,...,p_i$. The spectrum of the Hamiltonian (\ref{H1}) can be expressed in different ways depending on the choice of the quantum numbers.

\subsection{ The $N=3$ case}  \label{subs4}
To motivate our general discussions, in this subsection we examine the distinct subalgebra structures of ${\cal SW}(3)$. The Hamiltonian system (\ref{H1}) for $N=3$ case reads,
\begin{equation}
 H= -\frac{1}{2} (\partial_{1}^2 +\partial_{2}^2+\partial_{3}^2)+ b ( x_1^2+x_2^2+x_3^2) + \frac{a_1}{x_1^2}+\frac{a_2}{x_2^2}+\frac{a_3}{x_3^2},
\end{equation}
and the corresponding second order integrals of motion are  $B_1,B_2,B_3$ given in (\ref{thebs}) and $A_{12}, A_{13},A_{23}$ in (\ref{theas}).
They satisfy the following commutation relations, 
\begin{eqnarray}
&&[H,B_i]=0, \quad [H,A_{ij}]=0, \quad [B_i,B_j]=0, \quad i,j=1,2,3;
\nonumber\\&&
 [ A_{23},B_1]=0,\quad [ A_{13}, B_2]=0,\quad [A_{12}, B_3]=0. \label{E3.8}
\end{eqnarray}
 For later convenience, the diagrams below represent the above relations.
\begin{eqnarray*}
\begin{xy}
(0,0)*+{B_1}="m"; (20,0)*+{B_3}="p"; (40,0)*+{A_{12}}="n"; (20,20)*+{H}="r";   (60,0)*+{B_3}="j"; (80,0)*+{B_2}="i"; (100,0)*+{A_{13}}="l"; (80,20)*+{H}="g"; 
"j";"i"**\dir{--}; 
"l";"i"**\dir{--};
"i";"g"**\dir{--}; 
"l";"g"**\dir{--};
"g";"j"**\dir{--};
"r";"p"**\dir{--};
"m";"p"**\dir{--}; 
"n";"p"**\dir{--}; 
"n";"r"**\dir{--};
"r";"m"**\dir{--};
\end{xy}
\end{eqnarray*}
\begin{eqnarray}
\begin{xy}
(0,0)*+{B_2}="m"; (20,0)*+{B_1}="p"; (40,0)*+{A_{23}}="n"; (20,20)*+{H}="r";   
"r";"p"**\dir{--};
"m";"p"**\dir{--}; 
"n";"p"**\dir{--}; 
"n";"r"**\dir{--};
"r";"m"**\dir{--};
\end{xy} \label{Dia2}
\end{eqnarray}
We also have the following four linearly independent commutators of the second order integrals,
\begin{align}\nonumber
C_{12} &=[B_1, A_{12}]=-[A_{12}, B_2], \\
 C_{23}& =[B_2, A_{23}]=-[A_{23}, B_3],\nonumber\\  C_{31} &=[B_3, A_{31}]=-[A_{31}, B_1],\label{E3.9}
 \\
 D_{123}&=[A_{12},A_{31}]=[A_{13},A_{23}]=[A_{23},A_{12}]\nonumber. 
\end{align}
Then the diagram (\ref{Dia2}) shows that there are three possible subalgebras generated by three generators $\{A_{12}, B_1, C_{12}\}$, $\{A_{23}, B_2, C_{23}\}$  and $\{A_{31}, B_3, C_{31}\}$, respectively, which correspond to the following identification
\begin{eqnarray}
\{ E_{1}, F_1, C_1\} &\equiv & \{A_{12}, B_1, C_{12}\}, \nonumber\\
\{ E_{2}, F_2, C_2\} &\equiv & \{A_{23}, B_2, C_{23}\}, \nonumber\\
\{ E_{3}, F_3, C_3\} &\equiv & \{A_{31}, B_3, C_{31}\}. 
\end{eqnarray}
Each set satisfies the commutation relations (\ref{E19}) of the associated substructure with appropriate structure constants.

\subsection{The general $N$ case}
We now generalize the results in subsection \ref{subs4} to the general $N$ case and consider subalgebra structures generated by $\{B_i, B_j, A_{ij}; H, B_k, k=1,2,\dots, N \quad\text{and} \quad k\neq i,j\}$ for any fixed $i,j=1,2,\dots, N$.
By direct computations, we get the following quadratic subalgebra structure, denoted by ${\cal Q}_{ij}(3)$ for any fixed $i,j=1,2,\dots, N$,
\begin{align}
  [B_i, A_{ij}]&= C_{ij},              \nonumber \\
[B_i, C_{ij}]&=8B_i^2{-}8(2H{-}\sum_{k\neq i,j} B_{k})B_i{+}32b A_{ij} ,      \label{sal1}   \\
[A_{ij}, C_{ij}] &= -8\{B_i, A_{ij}\}{+}8(2H{-}{\sum_{k\neq i,j} }B_{k}) (A_{ij}+\sfrac{1}{2}[8a_i{-}3]){-}8(4a_i{+}4a_j{-}3)B_i. \nonumber
\end{align}
Comparing (\ref{sal1}) to (\ref{E19}), we have
\begin{eqnarray} \nonumber
&&\alpha_{ij} = 8,\quad \gamma_{ij} = 0,\quad \delta_{ij}= -8(2H{-}\sum_{k\neq i,j} B_{k}),\quad \epsilon_{ij}= 32b,\quad \zeta_{ij}=0, \\&&
a_{ij}=0, \quad d_{ij}=-8(4a_i{+}4a_j{-}3), \quad z_{ij}= 4(8a_i{-}3)(2H{-}\sum_{k\neq i,j} B_{k}).
\end{eqnarray}
The corresponding Casimir operator takes the form,
\begin{align}
K_{ij} &= C_{ij}^2-8\{B_i^2,A_{ij}\}+8(2H{-}\sum_{k\neq i,j} B_{k})\{B_i, A_{ij}\}-8 (4 a_i {+} 4 a_j{-}11)B_i^2\nonumber\\
& +8(8a_i{-}11)(2H{-}\sum_{k\neq i,j} B_{k})B_i-32b A_{ij}^2.
\end{align}
The Casimir operator can also be written in terms of only the central elements $H$ and all $B_k, k\neq i,j$ as
\begin{eqnarray}
K'_{ij} = 4(8a_i{-}3)\left(2H{-}\textstyle\sum_{k\neq i,j} B_{k} \right)^2{-}8b(8a_i{-}3)(8a_j{-}3).
\end{eqnarray}
In order to obtain the energy spectrum of the system (\ref{H1}) from the subalgebra (\ref{sal1}), we construct its realization in terms of the deformed oscillator algebra 
%\begin{eqnarray}
%&&\Phi(n_{ij}; u_{ij}, H)=\frac{1}{1024b^2}\left[4(n_{ij}{+}u_{ij}){-}2{-}2\nu_i\right]\left[4(n_{ij}{+}u_{ij}){-}2{+}2\nu_i\right]\nonumber\\&&\qquad\qquad \left[8b(n_{ij}{+}u_{ij}){-}4b{+}4b\nu_j{+}\sqrt{2b}(\sum_{k\neq i,j}B_k{-}2H)\right]\nonumber\\&&\qquad\qquad \left[8b(n_{ij}{+}u_{ij}){-}4b{-}4b\nu_j{+}\sqrt{2b}(\sum_{k\neq i,j}B_k{-}2H)\right].
%\end{eqnarray}
\begin{align}
&\Phi(n_{ij}; u_{ij}, H)=\frac{1}{1024b^2}\left[4(n_{ij}{+}u_{ij}){-}2{-}2\nu_i\right]\left[4(n_{ij}{+}u_{ij}){-}2{+}2\nu_i\right]\\ \nonumber
&\left[8b(n_{ij}{+}u_{ij}){-}4b{+}4b\nu_j{+}\sqrt{2b}(\sum_{k\neq i,j}B_k{-}2H)\right]\left[8b(n_{ij}{+}u_{ij}){-}4b{-}4b\nu_j{+}\sqrt{2b}(\sum_{k\neq i,j}B_k{-}2H)\right].
\end{align}
The values of parameter $u_{ij}$ and the eigenvalues of the operators $\sum_{k\neq i,j}B_k$ are determined by requiring that the corresponding representation of the deformed oscillator algebra (\ref{Deform}) is finite dimensional, i.e.,
\begin{eqnarray}
\Phi(p_{ij}{+}1;u_{ij},E)= 0, \quad \Phi(0;u_{ij},E)=0, \quad \Phi(n_{ij})>0, \quad\forall \quad  n_{ij}>0, \label{Const}
\end{eqnarray}
where $p_{ij}$ are positive integer. These constraints give 
\begin{align}
 u_{ij}&=\frac{1}{2}{+}\frac{\varepsilon_i \nu_i}{2}, \nonumber\\
 \sum_{k\neq i,j}B_k &= 2H-2\sqrt{2b}(p_{ij}{+}1{+}\varepsilon_i \nu_i{+}\varepsilon_j \nu_j),
\\ \nonumber
\Phi(n_{ij})&=n_{ij}(n_{ij}{+}\varepsilon_i \nu_i)(n_{ij}{-}p_{ij{}-}1)(n_{ij}{+}\varepsilon_j \nu_j{-}p_{ij}{-}1),
\end{align}
where $\varepsilon_i=\pm 1, \varepsilon_j=\pm 1$. The spectrum of $H$ could be determined in terms of the $p_{ij}$ where $i,j \in \{1,...,N\}$, from selected subsets of $N$ substructures. This can be seen alternatively guided by the form of the spectrum in Cartesian coordinates (\ref{En1}) and relation among the Hamiltonian and the $B_i$  operators (\ref{sumb}). Thus, we use the constraints (\ref{E3.22}) on the spectrum of $B_i$ which is given by
\begin{eqnarray}
e(B_i(x))=2\sqrt{2b}(2q_i+\varepsilon_i \nu_i+1),
\end{eqnarray}
and by virtue of (\ref{sumb}) the energy spectrum of system (\ref{H1}) is
\begin{eqnarray}
E =\sum_i^{N} \sqrt{2b}(2q_i+\varepsilon_i \nu_i+1). \label{En3}
\end{eqnarray}
Let us point out that the above results are obtained based on the algebraic manipulation only without using explicitly the corresponding Schr\"{o}dinger equation. In the next section we show that the Racah ${\cal R}( N)$ subalgebra can also be used to obtain the spectrum in the form (\ref{En2}).

\section{Algebraic derivation based on Racah algebra ${\cal R}({N})$ }

The Racah subalgebra is in fact related to the separation of variables in hyperspherical coordinates via the relation
\begin{eqnarray}
\sum_i \frac{\partial^2}{\partial x_i^2} =\frac{\partial^2}{\partial r^2}+\frac{N{-}1}{r}\frac{\partial}{\partial r}+ \frac{1}{r^2}\sum_{i<j}J_{ij}^2.
\end{eqnarray}
Let us define the new operator $Z$ associated with the separation of variables in hyperspherical coordinates,
\begin{eqnarray}
&&Z=\sum_{i<j}A_{ij}
=-\sum_{i<j}J_{ij}^2+2r^2\sum_i\frac{a_i}{x_i^2}-2\sum_i a_i+\frac{N(N{-}1)}{4},
\end{eqnarray}
such that (\ref{H1}) acquires the form
\begin{eqnarray}
H= -\frac{1}{2}\left[\frac{\partial^2}{\partial r^2} +\frac{N{-}1}{r} \frac{\partial}{\partial r}  -2 b r^{2}- \frac{4Z+8\sum_i a_i-N(N{-}1)}{4r^{2}}\right].\label{HZ}
\end{eqnarray}
Comparing the above equation with (\ref{An0}) and using (\ref{En2}) leads to the spectrum of $Z$,
\begin{align}
e(Z)&=k_1{-}2\sum_i a_i{+}\frac{N(N{-}1)}{4}\\
&= \left[2\sum_{i=1}^{N{-}1} \tau_{i}\pm \sum_{i=1}^{N} \nu_i {+}N{-}1\right]^2{-}\frac{1}{4}(N{-}2)^2{-}2\sum_i a_i {+}\frac{N(N{-}1)}{4}\\
&=\nu^2{+}\frac{1}{4}(3N{-}4){-}2\sum_i a_i,
\end{align}
where $\nu$ is given by (\ref{La1}) which allows to rewrite the eigenvalues (\ref{En2})  as 
\begin{equation}
E=\sqrt{2b}(2\tau_r+2\nu +1).\label{En4}
\end{equation}
The Hamiltonian written in terms of the radial variable $r$ and $Z$, where $Z$ can be seen as a Casimir of the Racah ${\cal R}(N)$ algebra, has similarities with the $N$-dimensional radial oscillator. This suggests the existence of ladder differential operators in the radial variable $r$,
\begin{equation}
\mathcal{D}^{\pm}=H\pm\sqrt{2b}r\frac{\partial}{\partial r}-2br^2\pm\sqrt{\frac{b}{2}}N,
\end{equation}
whose action on the wave functions is given by
\begin{equation}
\mathcal{D}^{+}\psi_{\tau_r}^{2\nu} =2\sqrt{2b}(\tau_r{+}1)\psi_{\tau_r+1}^{2\nu}, \quad \mathcal{D}^{-}\psi_{\tau_r}^{2\nu} =2\sqrt{2b}(\tau_r{+}2\nu)\psi_{\tau_r-1}^{2\nu}.
\end{equation}
Then $\mathcal{D}^{+}\mathcal{D}^{-}= 8b\tau_r(\tau_r{+}2\nu)\psi_{\tau_r}^{2\nu}$, $\mathcal{D}^{-}\mathcal{D}^{+}= 8b(\tau_r{+}1)(\tau_r+2\nu{+}1)\psi_{\tau_r}^{2\nu}$ and $\mathcal{D}^{\pm}$ satisfy the following $su(1,1)$ algebra relations,
\begin{eqnarray}
[\mathcal{D}^{+}, H]= -2\sqrt{2b}\mathcal{D}^{+},\quad [\mathcal{D}^{-}, H]= 2\sqrt{2b}\mathcal{D}^{-}, \quad [\mathcal{D}^{-}, \mathcal{D}^{+}]= 4\sqrt{2b}H.
\end{eqnarray}
This means that the spectrum of the $N$-dimensional SW system (\ref{H1}) can be obtained in same way as that for rotationally invariant systems with now the Racah algebra $\mathcal{R}(N)$  playing the same role as the angular momentum algebra. To show this, we define new integrals,
\begin{eqnarray}
&& Z_{l}=\sum_{1\leq i<k\leq l+1}A_{ik},\quad 1\leq l\leq N-2,\\&&
Y_{p}=\sum_{p\leq i<k\leq N}A_{ik},\quad 1\leq p\leq N-1.
\end{eqnarray}
We examine the subalgebra structures generated by $Y_i$, $Z_{i-1}$ and the central elements $Y_1, Y_{i+1}, Z_{i-2}$ with $2\leq i\leq N-1$, $Z_0= 0$ and $Y_N=0$. These elements obey the following quadratic algebra  relations,
\begin{align}
[Z_{i-1}, Y_i]&=C_i, \nonumber \\
[Z_{i-1}, C_i]&= 8Z_{i-1}^2+8\{Z_{i-1}, Y_i\} -8(Y_1{+}Y_{i+1}{+}Z_{i-2}{-}4a_i{+}3/2)Z_{i-1} \nonumber \\
&+4(8\sum_{j=1}^{i}a_j{-}3i)Y_i  -4(8a_i{-}3)Y_1 -4[8\sum_{j=1}^{i-1} a_j{-}3(i{-}1)]Y_{i+1}\nonumber \\
&+8(Y_1{-}Y_{i+1})Z_{i-2},  \label{Zal2}  \\
[Y_i, C_i]&= -8Y_i^2-8\{Z_{i-1}, Y_i\} -4[8\sum_{j=i}^{N} a_j{-}3(N{-}i{+}1)]Z_{i-1} \nonumber \\ &+8(Y_1{+}Y_{i+1}{+}Z_{i-2}{-}4a_i{+}3/2)Y_i+4(8a_i{-}3)Y_1 \nonumber \\
&+4[8\sum_{j=i+1}^{N}a_{j}{-}3(N{-}i)]Z_{i-2} -8Y_1Y_{i+1}+8Y_{i+1}Z_{i-2}.\nonumber
\end{align}
It follows that $\{Y_i, Z_{i-1}, C_i; Y_1, Y_{i+1}, Z_{i-2}, 2\leq i\leq N-1, Y_N\equiv 0, Z_0= 0\}$ form the subalgebra ${\cal R}(3)$ for fixed $i$. This quadratic subalgebra can be fitted into the general form (\ref{E19}) with 
\begin{eqnarray}\nonumber
&&\alpha_i = 8,\quad \gamma_i = 8,\quad \delta_i= -8(Y_1{+}Y_{i+1}{+}Z_{i-2}{-}4a_i{+}3/2),\quad \epsilon_i= 4(8\sum_{j=1}^{i}a_j{-}3i), \\&& \nonumber \zeta_i= -4(8a_i{-}3)Y_1 -4[8\sum_{j=1}^{i-1} a_j{-}3(i{-}1)]Y_{i+1} +8Y_1Z_{i-2}-8Y_{i+1}Z_{i-2}, \quad
a_i=0, \\&& d_i=-4[8\sum_{j=i}^{N} a_j{-}3(N{-}i{+}1)],\\&& \nonumber z_i= 4(8a_i{-}3)Y_1+4[8\sum_{j=i+1}^{N}a_{j}{-}3(N{-}i)]Z_{i-2}-8Y_1Y_{i+1}+8Y_{i+1}Z_{i-2}.
\end{eqnarray}
The corresponding Casimir operator involving only the central elements $Y_1, Y_{i+1}, Z_{i-2}$ takes the form 
\begin{eqnarray} \nonumber
&K_i'&= 4 ( 8 a_i{-}3)Y_1^2 - 64Y_1 Y_{i+1} + 4 [8 \sum_{j=1}^{i-1}a_{j}{-}3(i{-}1)]Y_{i+1}^2 + 32 (8 a_i{-}3)Y_1 \\&& \nonumber\quad -4( 8 a_i{-}3) [8 \sum_{j=1}^{i-1}a_{j}{-}3(i{-}1)]Y_{i+1} + 16 Z_{i-2}^2 Y_{i+1} - 64 Z_{i-2} Y_1 - 16 ( 8 a_i{-}3) Z_{i-2} Y_{i+1} \\&&\quad -4 ( 8 a_i{-}3)[8 \sum_{j=i+1}^{N}a_{j}{-}3(N{-}i)]Z_{i-2} + 4 [8 \sum_{j=i+1}^{N}a_{j}{-}3(N{-}i)]Z_{i-2}^2 \\&&  \nonumber\quad - 16 Z_{i-2} Y_1 Y_{i+1}  + 16 Z_{i-2} Y_{i+1}^2 -(8 a_i{-}3)[8 \sum_{j=1}^{i-1}a_{j}{-}3(i{-}1)] [8 \sum_{j=i+1}^{N}a_{j}{-}3(N{-}i)].
\end{eqnarray}
The subalgebra algebra (\ref{Zal2}) can be realized in terms of the deformed oscillator algebra (\ref{Deform}) with structure function given by 
\begin{eqnarray} \nonumber
\Phi(n_i,u_i)&=&[n_i{+}u_i{-}(2{-}y_1{-}y_{i+1})/4][n_i{+}u_i-(2{-}y_1{+}y_{i+1})/4]\\ \nonumber &&[n_i{+}u_i{-}(2{+}y_1{-}y_{i+1})/4][n_i{+}u_i{-}(2{+}y_1{+}y_{i+1})/4]\\&&[n_i{+}u_i-(2{+}z_{i-2}{+}2\nu_i)/4][n_i{+}u_i{-}(2{+}z_{i-2}{-}2\nu_i)/4]\\&& \nonumber [n_i{+}u_i{-}(2{-}z_{i-2}{-}2\nu_i)/4][n_i{+}u_i{-}(2{-}z_{i-2}{+}2\nu_i)/4],
\end{eqnarray}
where $y_1, y_{i+1}, z_{i-2}, \nu_i$ satisfy
\begin{align}
 Y_1&=\frac{1}{4} \left(3N{-}4{-}8\sum_{j=1}^N a_j {+}y_1^2\right),\nonumber\\
Y_{i+1}& = \frac{1}{4}  \left(3N{-}3i{-}4{-}8\sum_{j=i+1}^N a_j{+}y_{i+1}^2\right),\label{EY1}\\ \nonumber
Z_{i-2}&=\frac{1}{4}\left(3i{-}7{-}8\sum_{j=1}^{i-1} a_j{+}z_{i-2}^2\right).
\end{align}
Imposing the constraints $\phi(0,u_i)=0$ and $\phi(p_i+1,u_i)=0$ (which is the condition for the corresponding representation of the deformed oscillator algebra to be finite-dimensional), where $p_i$ is positive integer, we obtain
\begin{align}
u_{i}&=\frac{1}{4} (2 + \varepsilon_1 y_1 + \varepsilon_2 y_{i+1}),\quad \text{or}\quad u_{i}=\frac{1}{4} (2 + \varepsilon_1 z_{i-2} + 2\varepsilon_2 \nu_i),\\
z_{i-2}&=4\bar{\varepsilon}_1(p_i+1)+\bar{\varepsilon}_2 y_1 + \bar{\varepsilon}_3 y_{i+1}+2 \bar{\varepsilon}_4 \nu_i, \quad \text{or}\\
y_1&=4\bar{\varepsilon}_1(p_i+1) + \bar{\varepsilon}_2 y_{i+1}+2 \bar{\varepsilon}_3 \nu_i+\bar{\varepsilon}_4 z_{i-2},\label{rey1}
\end{align}
where $\varepsilon_1, \varepsilon_2, \bar{\varepsilon}_1, \bar{\varepsilon}_2, \bar{\varepsilon}_3, \bar{\varepsilon}_4$ take the values $\pm 1$.
In what follows we will take
\begin{eqnarray}
u_{i}=\frac{1}{4} (2 +  z_{i-2} + 2\nu_i).
\end{eqnarray}

Using the fact that the integrals $Z_{i-1}$ are diagonal in the number operator in the oscillator realization (\ref{E3.21}) and from (\ref{EY1}), we obtain
\begin{align}
Z_{i-1}&=\frac{\gamma}{2} \left[(q_i{+}u_i)^2{-}\frac{1}{4}{-}\frac{\epsilon}{\gamma^2}\right]=4(q_i{+}u_i)^2{+}\frac{1}{4}(3i{-}8\sum_{j=1}^i a_j){-}1, \\
&=\frac{1}{4}(3i{-}8\sum_{j=1}^i a_j){+}\frac{1}{4}z_{i-1}^2{-}1.
\end{align}
Comparing these two relations, we find that $z_{i-1}=4(q_i+u_i)$ and the recurrence relation,
\begin{equation}
z_{i-1}=4q_i+z_{i-2}+2\nu_i+2,
\end{equation}
with the initial condition $z_0=\nu_1$, giving as a result
\begin{equation}
z_{N-2}=4\sum_{i=1}^{N-1}q_i+2\sum_{i=1}^{N-1}\nu_i+2(N{-}2).\label{recu1}
\end{equation}
By virtue of (\ref{rey1}) and choosing suitable sign of $\bar{\varepsilon}_i$ and setting $y_{N+1}=0$, we can write
\begin{align}
y_1&=4\bar{\varepsilon}_1(p_N{+}1){+}\bar{\varepsilon}_2 y_{N+1}{+} 2\bar{\varepsilon}_3 \nu_N{+}\bar{\varepsilon}_4 z_{N-2}, \\
&=4(p_N{+}1){+}4\sum_{i=1}^{N-1}n_i{+}2\sum_{i=1}^{N}\nu_i+2(N{-}2).
\end{align}
The substitution of the above relation into the first equation in (\ref{EY1}) leads to
\begin{equation}
e(Z)=e(Y_1) 
= \left[2p_N+2\sum_{i=1}^{N-1} q_i +\sum_{i=1}^N \nu_i+N\right]^2+\frac{1}{4} (3N{-}4) - 2\sum_{j=1}^N a_j.
\end{equation}
To derive the spectrum of $H$, we consider the algebra generated by the integrals, $\{Y_1,B_N; H,Z_{N-2}\}$. It can be shown that these integrals close to form the following quadratic algebra 
\begin{align}
 [Y_1,B_N]&=D,\nonumber\\
[Y_1,D]&=8\{Y_1,B_N\}{-}16HY_1{+}4(8\sum_{j=1}^N a_j{-}3N)B_N{+}16HZ_{N-2}{-}8(8a_N{-}3)H, \label{Yal1} \\
[B_N,D]&=-8B_N^2{-}32bY_1{+}16HB_N{+}32bZ_{N-2}.\nonumber
\end{align}
Comparing (\ref{Yal1}) with (\ref{E19}), we have
\begin{eqnarray} \nonumber
&&\alpha = 0,\quad \gamma = 8,\quad \delta= -16H,\quad \epsilon= 4(8\sum_{j=1}^N a_j-3N), \\&& \zeta=  16HZ_{N-2}{-}8(8a_N{-}3)H, \quad
a=0, \quad d=-32b,\quad z= 32bZ_{N-2}.
\end{eqnarray}
The Casimir operator in terms of only the central elements $H$ and $Z_{N-2}$ has the form,
\begin{eqnarray}
 K'=32bZ_{N{-}2}^2{+}16(8a_N{-}3)H^2{-}32b(8a_N{-}3)Z_{N{-}2}{-}8b(8a_N{-}3)\left[8{\displaystyle \sum_{j=1}^{N-1}}a_j{-}3(N{-}1)\right].
\end{eqnarray}
The quadratic algebra (\ref{Yal1}) can be realized in terms of the oscillator algebra (\ref{Deform}) with the structure function,
\begin{align} \nonumber
\phi(n_N,u_N)&=\left[n_N{+}u_N{-}\left(2{-}\sqrt{2/b}H\right)/4\right]\left[n_N{+}u_N{-}\left(2{+}\sqrt{2/b}H\right)/4\right]\\& [n_N{+}u_N{-}(2{+}z_{N-2}{+}2\nu_N)/4][n_N{+}u_N{-}(2{+}z_{N-2}{-}2\nu_N)/4]\\& [n_N{+}u_N{-}(2{-}z_{N-2}{-}2\nu_N)/4][n_N{+}u_N{-}(2{-}z_{N-2}{+}2\nu_N)/4], \nonumber
\end{align}
where $ z_{N-2}$ satisfy
\begin{equation}
Z_{N-2}=\frac{1}{4}(3N{-}7{-}8\sum_{j=1}^{N-1} a_j {+}z_{N-2}^2).
\end{equation}
Imposing the constraints $\phi(0,u_N)=0$ and $\phi(p_N+1,u_N)=0$ (where $p_N$ is a positive integer) to the above structure function gives the solutions
\begin{align}
u_{N}&=\frac{1}{4} \left(2{+}\varepsilon_1 \sqrt{\frac{2}{b}}H\right),\quad \text{or}\quad u_{N}=\frac{1}{4} (2{+}\varepsilon_1 z_{N-2}{+}2 \varepsilon_2 \nu_N),\\
e(H)&=\sqrt{\frac{b}{2}}(4p_N{+}4{+}\varepsilon_1 z_{N-2}{+}2\varepsilon_2  \nu_N), \qquad\varepsilon_1,  \varepsilon_2=\pm 1.\\
Y_1&=\frac{\gamma}{2}\left[(n_N{+}u_N)^2{-}\frac{1}{4}{-}\frac{\epsilon}{\gamma^2 }\right].   
\end{align}
By means of the recurrence relation (\ref{recu1}), we have
\begin{equation}
e(H)=\sqrt{2b}\left(2p_N+2\sum_{i=1}^{N-1} q_{i} +\sum_{i=1}^{N} \nu_i+ N\right).
\end{equation}
This formula coincides with the result from separation of variables in hyperspherical coordinates (\ref{En2}).  Taking into account (\ref{En3}), this approach represent a second algebraic derivation based on set of commuting integrals. Of course, there are many distinct choices of set of commuting integrals and therefore other possible choices of substructures and providing the spectrum in terms of other quantum numbers. This emphasizes the fact that algebraic derivations of the spectrum for $N$-dimensional systems can be based only on differential operators and their operator algebra in different forms. More generically, their nature might be less straightforward than other connecting with known solutions obtained via different approaches.

\section{Conclusion}

The symmetry algebra of a $N$-dimensional quantum superintegrable system is in general a quite  complicated algebraic structure, and well beyond the scope of finite-dimensional Lie algebras. In this paper we have presented the complete symmetry algebra for the $N$-dimensional quantum Smorodinsky-Winternitz system. It is a higher rank quadratic algebra, referred to as ${\cal SW}(N)$, and contains the Racah algebra ${\cal R}(N)$ as a subalgebra.

Provided the algebra of the conserved quantities, we have addressed the problem of algebraically determining the spectrum. Although the Smorodinsky-Winternitz superintegrable system was studied by several authors over the years from different perspectives such as separation of variables, the derivation of its spectrum via the underlying symmetry algebra remained until now as an open problem. This method is particularly relevant as for many of the superintegrable Hamiltonians, their potentials are related with nonlinear differential equation, i.e. Painlev\'e trascendents, so the standard methods of ordinary differential equation can not be directly applied. This fact supports the need of developing algebraic methods relying only on the explicit differential operators form. Those methods have, however, limitations as they do not take into account the explicit physical states and therefore boundary conditions or singularities. 

The two distinct approaches discussed here rely on the construction of different sets of substructures involving three generators (and central elements). For each of the substructures, the corresponding deformed oscillator algebra and their cubic Casimir are exploited. This demonstrates that the algebraic derivation is not unique for a superintegrable system. In one of the constructions, the subalgebra ${\cal R}(N)$ has been used, which points to the usefulness of the Racah algebra in a wider class of superintegrable systems beyond the usual models on a sphere. Finally, as we demonstrated, the higher rank quadratic algebras are useful in deriving the spectrum of a Hamiltonian in quantum mechanics. However,  they are also interesting mathematical objects by their own, which complete classification and further properties regarding representation theory are far to be complete.

\section*{Acknowledgement}
FC was supported by Fondecyt grants 1171475 \& 1211356. FH was supported by University Grant Commission, Bangladesh. IM was supported by Australian Research Council Future Fellowship FT180100099. YZZ was supported by Australian Research Council Discovery Project DP190101529 and National Natural Science Foundation of China (Grant No. 11775177).

%\newpage

\end{document}